\documentclass[aps,prl,reprint,groupedaddress]{revtex4-1}
\usepackage{graphicx}
\usepackage{color}

\usepackage[normalem]{ulem}

\begin{document}

%\preprint{}

\title{Spin-orbit splitting and effective masses in p-type GaAs two-dimensional hole gases}


\author{Fabrizio Nichele}
\email[]{fnichele@phys.ethz.ch}
\homepage[]{www.nanophys.ethz.ch}
\affiliation{Solid State Physics Laboratory, ETH Z\"{u}rich, 8093 Z\"{u}rich, Switzerland}

\author{Atindra Nath Pal}
\affiliation{Solid State Physics Laboratory, ETH Z\"{u}rich, 8093 Z\"{u}rich, Switzerland}

\author{Roland Winkler}
\affiliation{Department of Physics, Northern Illinois University, DeKalb, Illinois 60115, USA}

\author{Christian Gerl}
\affiliation{Universit\"{a}t Regensburg, Universit\"{a}tsstrasse 31, 93053 Regensburg, Germany}

\author{Werner Wegscheider}
\affiliation{Solid State Physics Laboratory, ETH Z\"{u}rich, 8093 Z\"{u}rich, Switzerland}

\author{Thomas Ihn}
\affiliation{Solid State Physics Laboratory, ETH Z\"{u}rich, 8093 Z\"{u}rich, Switzerland}

\author{Klaus Ensslin}
\affiliation{Solid State Physics Laboratory, ETH Z\"{u}rich, 8093 Z\"{u}rich, Switzerland}


\date{\today}

\begin{abstract}
We present magnetotransport measurements performed on two-dimensional hole gases embedded in carbon doped p-type GaAs/AlGaAs heterostructures grown on [001] oriented substrates. A pronounced beating pattern in the Shubnikov-de Haas oscillations proves the presence of strong spin-orbit interaction in the device under study. We estimate the effective masses of spin-orbit split subbands by measuring the temperature dependence of the Shubnikov-de Haas oscillations at different hole densities. While the lighter heavy-hole effective mass is not energy dependent, the heavier heavy-hole effective mass has a prominent energy dependence, indicating a strong spin-orbit induced non parabolicity of the valence band. The measured effective masses show qualitative agreement with self-consistent numerical calculations.
\end{abstract}

%\keywords{}

\maketitle


The understanding of any semiconductor material starts with the knowledge of the carriers effective mass and its energy dependence. For the most important semiconductors, such as Si and GaAs, the electron effective mass has been widely investigated using temperature dependent transport and cyclotron resonance experiments \cite{Dresselhaus1955, Smith1972, Shoenberg1984, Coleridge1989, Fletcher1990, Coleridge1996, Pan1999}. For two-dimensional hole gases (2DHG) in GaAs the situation is significantly more complicated. Despite the importance of GaAs for fundamental research and technological applications, a detailed study of the effective mass of holes in GaAs 2DHG grown along the high symmetry [001] direction remains to be done. The interpretation of the rapidly increasing number of experiments performed in 2DHGs requires a solid understanding of the physics underlying the effective mass value and its dependence on quantities such as hole density and spin-orbit interaction (SOI) strength.

Holes in the valence band of GaAs are characterized by wave functions whose symmetry in real space is reminiscent of atomic p-orbitals. Due to the interplay of the non-zero orbital angular momentum, bulk SOI and confinement in growth direction, the carriers in 2DHGs are effectively described as heavy holes with total angular momentum $z$ component $\pm 3/2$, for which SOI corrections are expected to be stronger than for their spin-1/2 electronic counterparts. SOI breaks $\pm 3/2$ total angular momentum degeneracy already at zero magnetic field, resulting in a band-warping. Accordingly, spin and momentum eigenstates mix, leading to a profound difference between the two spin-orbit-split (SO-split) bands \cite{Yu2003}. In p-type 2DHG, the main contribution to SOI is of Rashba type and originates from the structure inversion asymmetry of the host heterostrucure. Unlike the case of electrons, Rashba SOI for holes is expected to have a cubic dependence on the in-plane momentum \cite{Winkler2003}. With respect to other materials, GaAs 2DHGs offer the unique opportunity to study pronounced SOI effects in a system that can be grown with high control \cite{Noh2003, Huang2006} and reliably processed into nanostructures \cite{Grbic2007, Quay2010, Komijani2010, Srinivasan2012, Nichele2013, Komijani2013}. Furthermore, holes in GaAs have a theoretically predicted effective mass several times larger than electrons in the conduction band. The smaller Fermi energy makes the carrier-carrier Coulomb interactions more relevant, allowing the study of many-body related effects \cite{Noh2003, Huang2006, Komijani2010}.

\begin{figure*}
\includegraphics[width=\textwidth]{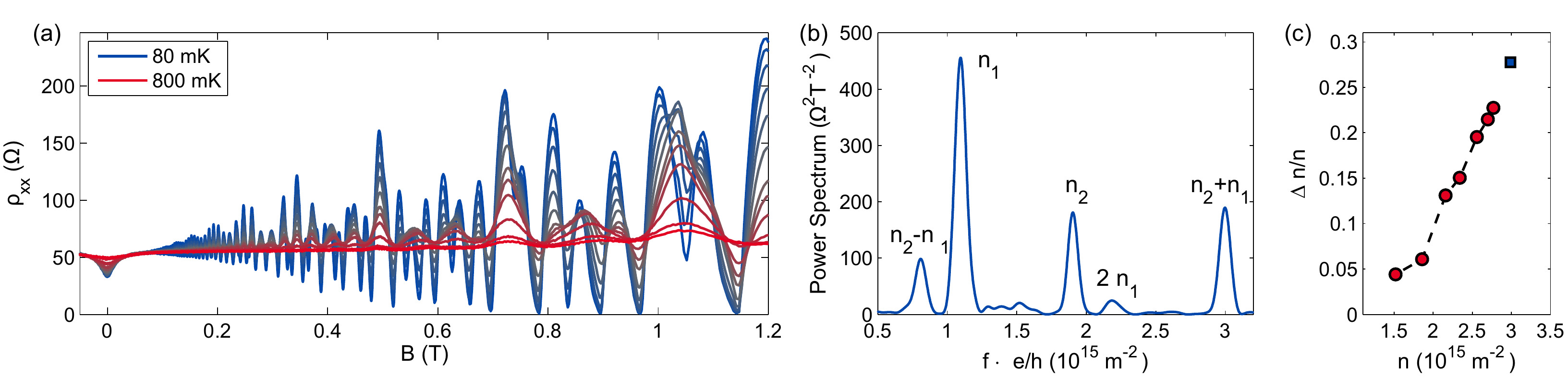}
\caption{(color online). (a) Longitudinal resistivity of the
ungated sample for various temperatures, from $80~\rm{mK}$ (blue line) to $800~\rm{mK}$ (red line). (b) Power spectrum of the low temperature magnetoresistance (as a function of $1/B$) shown in (a). (c) Zero-field SO splitting as a function of carrier density for the ungated (square) and gated (dots) sample.}
\label{fig1}
\end{figure*}

The strong SO-splitting in 2DHGs can be observed from the presence of a beating in the low-field Shubnikov-de Haas (SdH) oscillations \cite{Stormer1983, Eisenstein1984, Grbic2004, Habib2004, Grbic2008, Habib2009}. In an approximate picture, the beating is due to the presence of different sets of SdH oscillations for the two angular momentum eigenstates (referred to as 1 and 2), that contribute to transport in parallel. Each set $i$ is characterized by a density $n_i$, an effective mass $m_i$, a Drude scattering time $\tau_i$ and a quantum scattering time $\tau_{qi}$. Upon performing a Fourier transform of the longitudinal resistivity as a function of $1/B$, two peaks corresponding to the two subband densities are observed. The frequency axis $f$ can be directly mapped into densities $n$ by $n=fe/h$. Since they are coupled by SOI, and since scattering and charge redistribution between subbands can be present, various non-linear terms are expected \cite{Alexandrov1996}.

The carriers' effective mass $m^*$ in a two-dimensional system can be estimated from the temperature dependence of the low-field Shubnikov-de Haas (SdH) oscillations. Based on the Ando formula for single sub-band systems \cite{Ando1982}, the relative amplitude decay ${\Delta \rho_{xx}}/\rho_{xx}$ of the oscillations of the longitudinal resistivity $\rho_{xx}$ at a magnetic field $B$ can be fitted with the equation \cite{Ihn2010}:

\begin{equation}
\label{eq1}
\frac{\Delta\rho_{xx}}{\rho_{xx}}= 2~\exp\left(-\frac{\pi}{\omega_c\tau_q}\right)~\frac{2\pi^2k_BT/\hbar\omega_c}{\sinh\left(2\pi^2k_BT/\hbar\omega_c\right)} \; ,
\end{equation}

where $T$ is the temperature and $\omega_c=eB/m^*$ the cyclotron frequency. The fitting parameters are $\tau_q$ and $m^*$. The presence of two sets of SdH oscillations due to the two subbands makes it difficult to extract the two effective masses separately. If the magnetic field onsets of the oscillation differ, one of the two effective masses can be deduced from the $\rho_{xx}$ oscillations where the contribution of only one subband is relevant. The other effective mass can then be inferred assuming parabolic bands, hence $m_1/m_2=n_2/n_1$ as in Ref.~\onlinecite{Eisenstein1984}, or assuming $m_1/m_2=({\tau_2}/{\tau_1})$ as in Ref.~\onlinecite {Grbic2004}. In Ref.~\onlinecite{Habib2004, Habib2009} a filtering technique was used to separate the different contributions in Fourier space, yielding the individual masses without further assumptions.
Despite substantial differences in the effective mass values reported by previous works, the low density subband was always assigned a lower effective mass than the high-density subband. Therefore, the low density SO-split subband is referred to as light-heavy-hole (HHl) subband and the high density one as heavy-heavy-hole (HHh) subband. In Ref.~\onlinecite{Eisenstein1984} and \onlinecite{Habib2004, Habib2009} a linear dependence of the effective masses with respect to magnetic field was observed. The origin of the magnetic field dependence remained unclear and the limited density tunability did not allow a density dependent investigation. We report here accurate measurements of the effective masses $m_1$ and $m_2$ of the two SO-split $\pm 3/2$ sub-bands in the limit of small magnetic fields. A pronounced difference between $m_1$ and $m_2$ (up to a factor of three) and the absence of any field dependence is observed. While the HHl effective mass is found to be independent of density, the HHh effective mass shows a strong density dependence.

The wafer structure used for this experiment was grown by molecular beam epitaxy on a [001] oriented GaAs substrate. From the top surface, it consists of a $5~\rm{nm}$ GaAs capping layer, a $15~\rm{nm}$ AlGaAs layer homogeneously doped with carbon, a $25~\rm{nm}$ AlGaAs spacer and a $15~\rm{nm}$ wide GaAs quantum well. The asymmetric doping scheme creates a strong structural inversion asymmetry, so the holes' wavefunction mainly resides on the top side of the GaAs quantum well. From this wafer two samples were processed, each consisting of two $50~\rm{\mu m} \times 25~\rm{\mu m}$ Hall bars oriented perpendicularly to each other. The Hall bar structures were obtained by standard photolithography and chemical wet etching. One sample was covered by a $200~\rm{nm}$ $\rm{Si_3N_4}$ layer grown by plasma enhanced chemical vapor deposition and a Ti/Au global topgate deposited by shadow mask evaporation. The ungated sample showed a density of $3.0 \times 10^{15}~\rm{m}^{-2}$ and a mobility of $65~\rm{m^2V^{-1}s^{-1}}$. The presence of the gate insulator decreases the hole density to $2.1 \times 10^{15}~\rm{m}^{-2}$, the application of a top gate voltage allowed tuning the density from $2.8 \times 10^{15}~\rm{m}^{-2}$ to less than $1.0 \times 10^{15}~\rm{m}^{-2}$. No dependence of the measured quantities was observed for the two different Hall bar directions. The two samples were measured in $^3\rm{He}/^4{He}$ dilution refrigerator with a base temperature of $80~\rm{mK}$ using standard low frequency lock-in techniques. Currents below $10~\rm{nA}$ were used to avoid heating effects.

Fig. \ref{fig1}(a) shows the longitudinal resistivity measured in the ungated sample as a function of magnetic field and temperature. At base temperature (blue line), $\rho_{xx}$ shows a beating pattern in the SdH oscillations while at $800~\rm{mK}$ (red line) many SdH minima are completely suppressed and the remaining oscillations have a regular structure with clear $1/B$ periodicity. Fig. \ref{fig1}(b) shows the power spectrum of $\rho_{xx}$ at $80~\rm{mK}$ transformed as a function of $1/B$. The peaks corresponding to the HHl and HHh subbands are marked as $n_1$ and $n_2$ respectively. The $n_1$ peak is directly assigned since its frequency corresponds to the low-field periodicity of the SdH oscillations. The peak labeled $n_1+n_2$ accurately matches the total density derived from the Hall effect. The difference in frequency between the $n_1$ peak and the $n_1+n_2$ peak allows to identify the second subband peak, labeled $n_2$. The peak labeled $n_2-n_1$ matches the difference between the subband densities while the $2n_1$ peak is a second harmonic of the $n_1$ peak. The positive magnetoresistance visible for a magnetic field smaller than $100~\rm{mT}$ is understood in terms of classical two-band transport \cite{Zaremba1992,Grbic2008} and constitutes further evidence of strong SOI. For this kind of analysis it is common to multiply the data with a smooth windowing function to suppress the boundary effect in the final results. A detailed description of the numerical procedure used to transform the data can be found in the supplementary material. The SO-splitting, quantified here as $\Delta N /N =(n_2-n_1)/(n_2+n_1)$ varies with gate voltage \cite{Lu1998, Grbic2008}. The density dependence of the SO-splitting is shown in Fig. \ref{fig1}(c) for the ungated (blue square) and the gated device (red dots). An estimation of the spin-orbit energy splitting between subbands can be found in the supplementary material.

\begin{figure}
\includegraphics[width=\columnwidth]{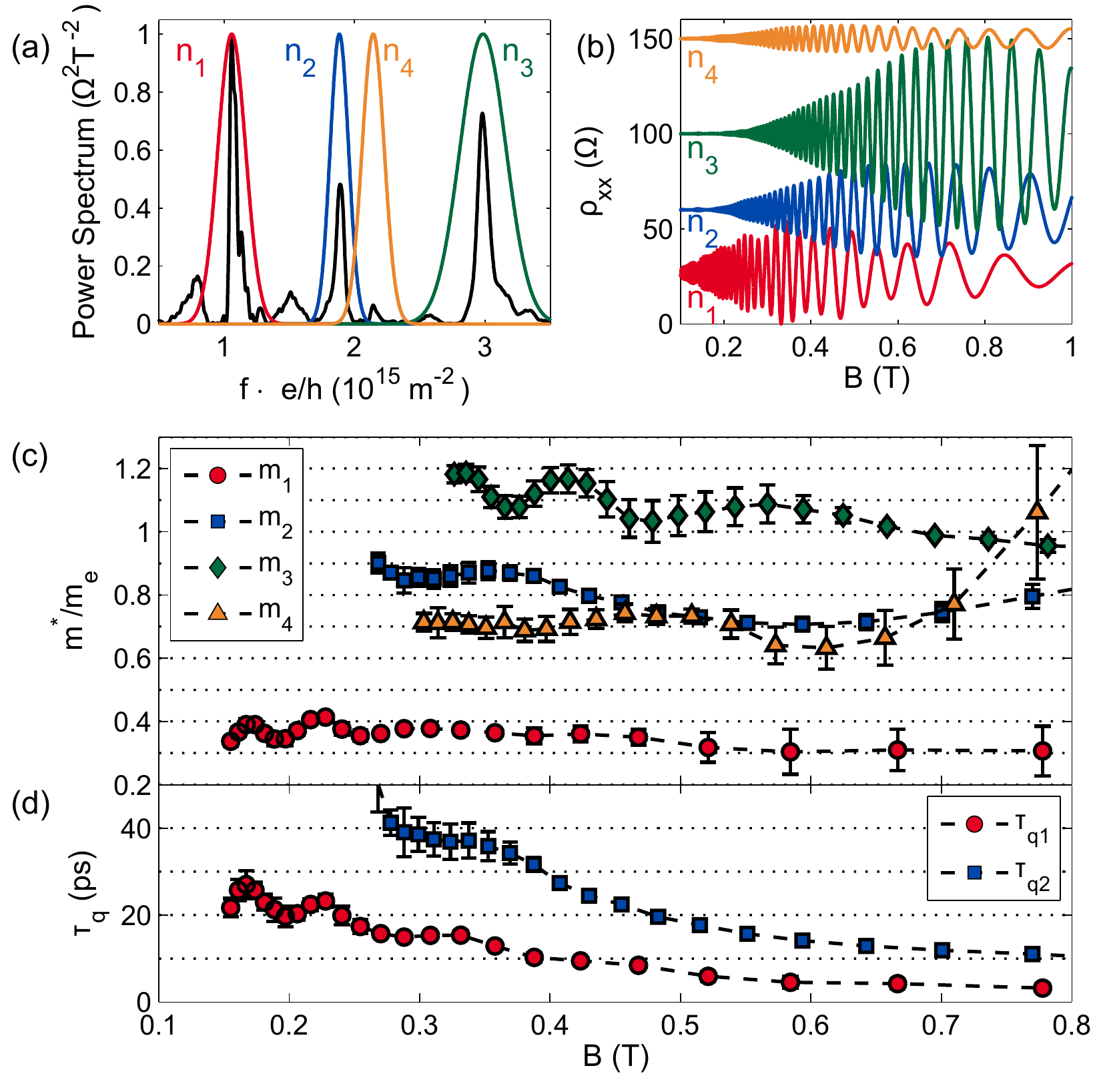}
\caption{(color online). Analysis using Method A. (a) Power spectrum of the low temperature magnetoresistance at density $n = 3.0 \times 10^{15}$~m$^{-2}$ together with the filters used to extract the different components. The power spectrum has been normalized in order to compare it with the filters. No windowing is performed. (b) SdH oscillations obtained after inverse Fourier transforming the filtered spectrum, the oscillations have been vertically offset for clarity. (c) Effective masses deduced from the filtered oscillations as a function of magnetic field. $m_1$ and $m_2$ are the effective masses of the two SO-split subbands, $m_3$ and $m_4$ are fictitious effective masses describing the temperature dependences of the $n_2+n_1$ and $2n_1$ peak respectively. (d) Quantum scattering times of the two SO-split subbands.}
\label{fig2}
\end{figure}

We used two distinct methods to extract the effective masses from the temperature dependence of the SdH oscillations, referred to as Methods A and B. Method A is adapted from Ref.~\onlinecite{Habib2004, Habib2009} and consists of separating the different spectral components by finite-width spectral filters. Once a peak is isolated, its inverse Fourier transform reveals the corresponding SdH oscillations. The isolated oscillations are added to the slowly varying background, obtained by fitting $\rho_{xx}$ to a low-order polynomial, and the standard procedure to extract the effective mass is applied to the newly obtained data. Windowing the raw data set should be avoided here, since it can substantially modify the amplitude of different frequency components. The presented data are obtained using Gaussian windows as filters. The width of each filter is chosen to be as large as possible, to avoid both perturbing the shape of the peak and including spurious frequency components in the filtered data. We checked that the final results are independent of the particular filter shape and robust against moderate modification of the filter width. For very small magnetic field, due to  limited oscillation amplitude, we could not satisfactorily fit the model to the data, hence those points were excluded from the analysis.  Fig. \ref{fig2}(a) shows the filters used for analyzing the data of Fig. \ref{fig1} and Fig. \ref{fig2}(b) gives the corresponding SdH oscillations. Fig. \ref{fig2}(c) shows the effective masses obtained by fitting Eq. (\ref{eq1}) to the minima of the filtered oscillations and Fig. \ref{fig2}(d) the quantum scattering times obtained for $n_1$ and $n_2$. In contrast to previous works we clearly see that, in the limit of small magnetic field, the effective masses $m_1$ and $m_2$ do not depend on $B$. As the magnetic field increases beyond about $350~\rm{mT}$ we leave the validity range of Eq.\ (\ref{eq1}) since the oscillations' amplitude becomes comparable to the background level (about $50~\rm{\Omega}$). Here, any analysis based on Eq.\ (\ref{eq1}) should be avoided. Alternatively, the magnetic field in which the amplitude of the $n_2+n_1$ oscillations becomes relevant (about $350~\rm{mT}$), can be used as the limit for the validity range of the analysis. From the data points at low magnetic field we estimate $m_1=\left(0.374\pm0.003\right)m_e$ and $m_2=\left(0.88\pm0.01\right)m_e$, $m_e$ being the free electron mass and $\tau_{q1}=23.4\pm0.8~\rm{ps}$ and $\tau_{q2}=39\pm1.5~\rm{ps}$. The quantum scattering times are an order of magnitude lower than the Drude scattering times obtained from the classical positive magnetoresistance. The oscillations in $m_1$ and $\tau_{q1}$ visible at small magnetic field are due to side peaks in the power spectrum in Fig. \ref{fig2}(a). They originate from boundary effects in the Fourier transform and are totally suppressed by windowing the data, as shown in Fig. \ref{fig1}(b).  We further investigated the temperature dependence of  the $n_2+n_1$ and $2n_1$ peaks, assigning them fictitious effective masses $m_3$ and $m_4$ respectively. The $2n_1$ peak is the second harmonic of $n_1$. As expected, an analysis based on Eq.\ (\ref{eq1}) gives an effective mass of $2m_1$ \cite{Ihn2010}. The $n_2+n_1$ peak has the strongest temperature dependence found, compatible with an effective mass of $m_1+m_2$.  The analysis could not be performed on other peaks due to their strong temperature dependence and small amplitude. In particular the $n_2-n_1$ peak cannot be easily filtered from the low frequency background relevant at high temperature.  Qualitatively similar results were obtained with the gated sample for densities larger than $2.5 \times 10^{15}~\rm{m}^{-2}$. The analysis was not possible for smaller densities since the decrease in $\tau_q$ and SO splitting make the separation between peaks too small to apply sufficiently broad filters and avoid overlaps.

\begin{figure}
\includegraphics[width=\columnwidth]{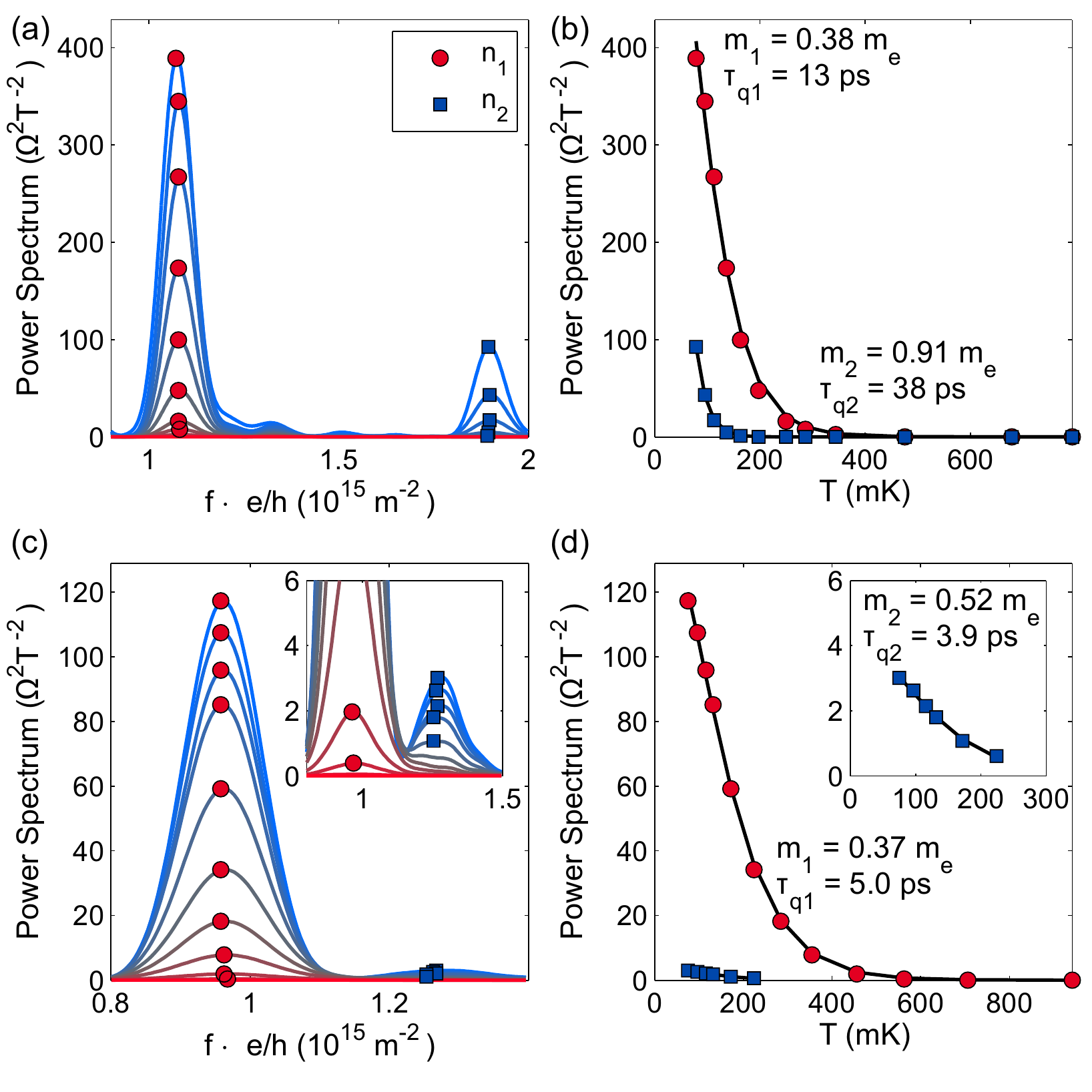}
\caption{(color online). Analysis using Method B. (a) Temperature dependence of the resistivity power spectrum in the ungated device. Dots and squares indicate the height of the $n_1$ and $n_2$ peak respectively. (b) Peaks heights as a function of temperature together with a fit (line). (c) and (d) The same as in (a) and (b) for the gated sample. The density was $2.1\times10^{-15}~\rm{m^{-2}}$ and the SO splitting $13\%$. The insets are zoom-ins of the $n_2$ peak.}
\label{fig3}
\end{figure}

The second method, called Method B, relies on the temperature decay of the peaks in the power spectrum. Given a magnetic field interval, one can numerically construct $\rho_{xx}(B)$ from the Ando formula \cite{Ando1982} and Fourier transform it in order to compare the peak height with the measured data. The zero-field resistivity and the hole density are read from the experimental data while $m^*$ and $\tau_q$ are fitting parameters. To provide robustness to the procedure, the fit is performed on the amplitude of a peak as a function of temperature. Method A requires the definition of a functional form for the filters, and cannot be applied for small separation between successive peaks. Method B only requires the input of a magnetic field range and does not use any finite-width filters. It can thus be applied to situations with limited SO splitting. Furthermore, any additional modification of the data set (e.g. windowing) can be implemented without side effects as long as it is identically applied to both the experimental and the calculated resistivities. Fig. \ref{fig3} shows the procedure for the two extreme cases where the method was applied. On the left side we see how the $n_1$ and $n_2$ peaks decay with temperature, on the right the peak amplitudes (markers) are fitted to the numerical model (solid lines). The results are indicated in the figure (errorbars are within $\pm5\%$), and are compatible with the quantitative findings of Method A. In the limit of small magnetic field, the obtained results do not show any dependence on the specific magnetic field windows chosen for the analysis.

\begin{figure}
\includegraphics[width=\columnwidth]{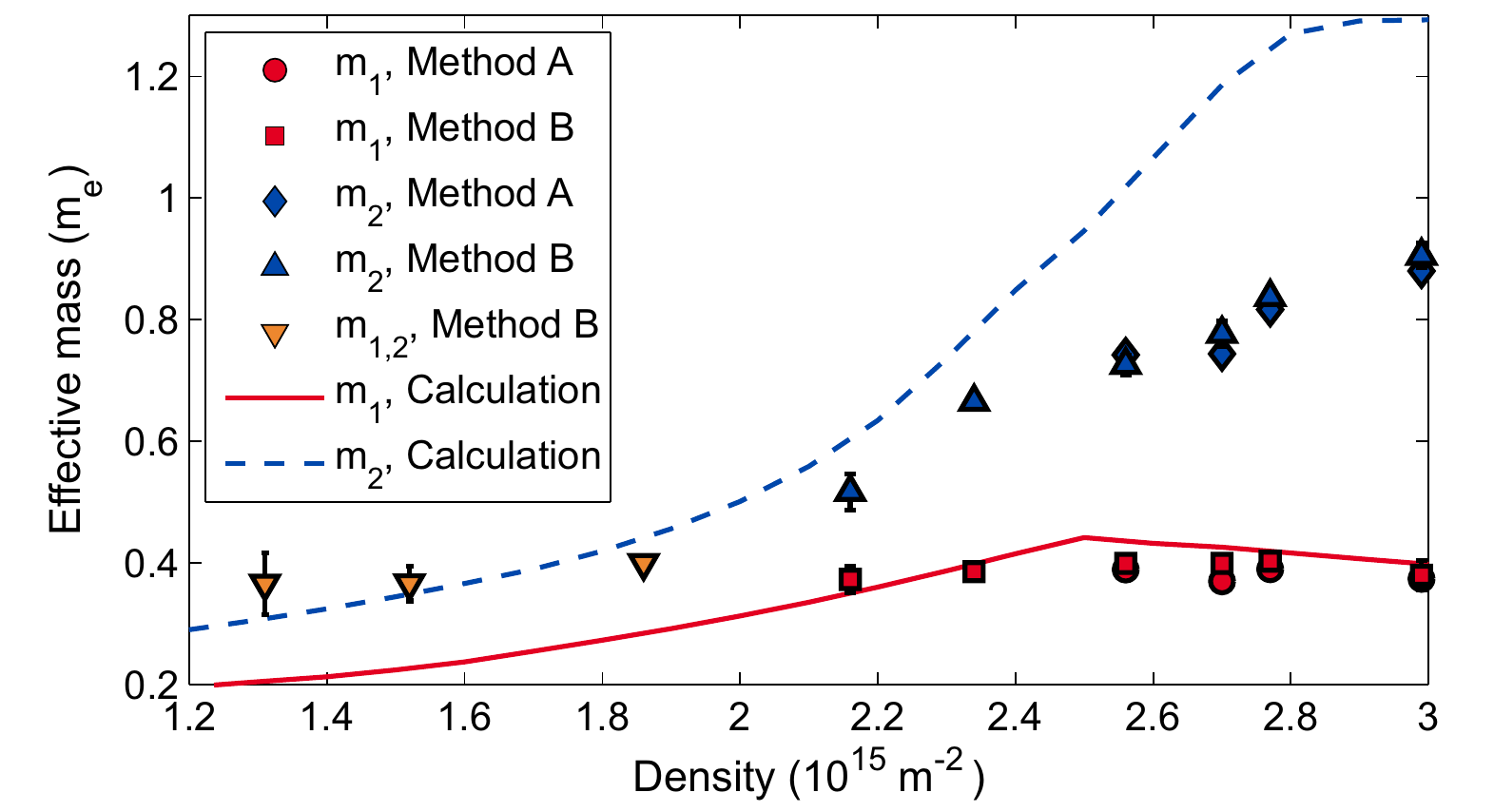}
\caption{(color online). Effective masses as a function of density. Comparison of the results for $m_1$, $m_2$ and $m_{1,2}$ obtained using Method A, Method B and self-consistent calculations.}
\label{fig4}
\end{figure}

Fig. \ref{fig4} summarizes the result of our analysis. Both methods proposed here can be applied to obtain the two different effective masses when a clear SO-splitting is present, so for sufficiently high hole densities. Method A requires a larger SO-splitting than Method B, so data points are provided only for higher densities. When both methods are applicable, the obtained results nicely match providing consistency for the analysis performed. At low density only one peak is visible in the spectrum, hence only one effective mass $m_{1,2}$ is resolved. The HHl effective mass is constant within the density range under study and equal to $0.38~m_e$. The HHh effective mass is instead strongly dependent on the carrier density, indicating a SOI induced non-parabolicity of the valence band, with a less than parabolic dependence on $k$. Both methods precisely determine the fitting parameters. The error bars reported here only refer to statistical errors, and are comparable to the estimated systematic errors in the measurements, e.g. a possible calibration error of the temperature read-out.

The experimental findings are in good agreement with theoretical predictions on the density dependence of the SO-split density-of-states effective masses at the Fermi energy in the limit $B \rightarrow 0$ of a GaAs 2DHG grown on the [001] surface. In our self-consistent calculations we used the slope of the Hartree potential at the back interface of the quantum well as a fitting parameter to reproduce the SO-splitting measured for the density of $3.0\times 10^{15}~\rm{m^{-2}}$. This slope was then kept fixed when modeling the different densities tuned via a topgate. The final results are shown in Fig. \ref{fig4} (solid lines). The calculated effective masses obtained in the limit $B \rightarrow 0$ show good agreement with the low-field experimental results both in terms of magnitude and trends. Caution should be paid when quantitatively comparing experimental results with self-consistent calculations. Different effective mass definitions can give rise to pronounced differences in the calculated results for a material system with strong band non-parabolocities and high anisotropies such as p-type GaAs. We remark that different experimental techniques or theoretical approaches give access to different properties of the system and could therefore result in slightly different effective mass values.

In conclusion, we extracted the effective masses of SO-split subbands in p-type 2DHGs grown along the [001] direction. Two different methods allow us to obtain the two effective masses separately. The high quality of our samples allows us to measure at very low magnetic field, where Eq. (\ref{eq1}) is valid, and rule out the linear dependence of the effective mass on magnetic field observed in previous work. In the accessible density range the HHl effective mass is constant, the HHh effective mass shows a strong density dependence due to SOI induced non-parabolicities in the valence band. The experimental results are qualitatively confirmed by self-consistent calculations. The effective masses in hole systems are markedly different for the two SO-split subbands and strongly dependent on sample specific properties such as density and SOI strength. These results highlight the complexity of the valence band of GaAs.

\begin{acknowledgments}
  The authors wish to thank Yashar Komijani and Szymon Hennel for useful discussions and the Swiss National Science Foundation for financial support.
\end{acknowledgments}

\bibliography{Bibliography}

\end{document}